\documentclass[12pt]{article}
\usepackage[top=1in,bottom=1in,left=1in,right=1in]{geometry}
\usepackage{authblk}
\usepackage{blindtext}
\usepackage{amsmath,amssymb,amsthm}
\usepackage[T1]{fontenc}
\usepackage[tt=false]{libertine}
\usepackage[scaled=0.83]{beramono}
\usepackage[libertine]{newtxmath}
\usepackage{graphicx}
\usepackage{bm}
\usepackage{array,multirow}
\usepackage{booktabs}
\usepackage{mathtools}
\mathtoolsset{showonlyrefs}
\usepackage{definitions}

\allowdisplaybreaks

\begin{document}

\title{Mermin Devices for Generalized Dicke States}

\author[1]{Roman V. Buniy\thanks{roman.buniy@gmail.com}}
\author[2]{Thomas W. Kephart\thanks{tom.kephart@gmail.com}}
\affil[1]{Schmid College of Science, Chapman University, Orange, CA 92866}
\affil[2]{Department of Physics and Astronomy, Vanderbilt University, Nashville, TN 37235}

\date{\today}

\maketitle

\begin{abstract}
    We present here several new exact results for a number of entangled states: the W-state of three qubits and its generalization --- Dicke states for more than three qubits.
    We derive these results by bounding the expected values of the Bell-Mermin operators.
    We review the three qubit GHZ Mermin device, make its generalization to four qubits, and then construct analogous Mermin devices for the generalized Dicke states of three and four qubits.
    As a result of studying if their operations can be fully explained by Mermin's instructional sets, we show that the GHZ and Dicke states of three qubits and the GHZ state of four qubits do not allow such a description.
    However, among the two generalized Dicke states of four qubits, one does allow and the other does not allow such a description.
\end{abstract}
 
\section{Introduction}
\label{section_introduction}

It is well known that entanglement can be considered as a resource when characterizing quantum states. 
The full description of this resource and its resulting practical uses, however, is an open problem in quantum information science.
Studying inequalities of the Bell type is the first step towards solving this problem as they provide bounds on the amount of information stored in entangled states.
The exact results for such bounds are known only for a very limited set of states.

In terms of entanglement, the Bell inequalities \cite{Bell:1964kc,Bell:1964fg} and further 
inequalities studied by  Clauser, Horne, Shimony and Holt (CHSH) \cite{Clauser:1969ny} and Tsirelson \cite{Tsirelson} relate noncommutativity in physical systems to quantum entanglement as a resource.
Perhaps Mermin illuminated this topic the most clearly in his papers \cite{Mermin:1990AJP,Mermin:1990vxe,Mermin:1990dqo,Mermin:1993zz}, where he introduced what are now called Bell-Mermin operators and used them to explore how entanglement is associated with the eigenvalues of these operators. 

Mermin focused on the three qubit GHZ state \cite{Greenberger:1989tfe} and its generalization to multi-partite qubits, and obtained exact results for the violation of Bell type inequalities.
Besides the GHZ state, the W state \cite{Dur:2000zz} is the only other fully entangled three qubit state, and numerical results are known for its violation of the inequalities.
There is not a lot more in the literature about the violation of Bell type inequalities of four or more qubits and it is our purpose here to begin to provide such information as well as to give some new precise results for three qubits. 

The degree to which a state can violate Bell type inequalities gives us some notion of how useful it can be in a quantum computer or other quantum device.
Below we first present exact analytical results for the three qubit W state.
We also give  analytic results for a few other cases, as well as numerical results for a number of interesting cases including generalized W states, which are a subset of the Dicke states \cite{Dicke}.
As both W and their Dicke state generalizations retain some amount of quantum entanglement even when one or more  qubits are traced out, they have a degree of robustness not shared with GHZ states, hence we expect the study of these and other classes of multi-qubit states to provide a broad range of parameter space for use in quantum devices.
Our focus is on pure states throughout, but the results could be extended to mixed states, if required.
However, we show that there is still much to learn about pure state entanglement. 

\section{States}
\label{section_states}

We first define several relevant states of $n$ qubits.

We choose an orthonormal basis $\{e_{1},e_{2}\}=\bigl\{\begin{psmallmatrix} 1 \\ 0 \end{psmallmatrix},\begin{psmallmatrix} 0 \\ 1 \end{psmallmatrix}\bigr\}$ of the vector space $V$ of each qubit, and set tensor products $e_{j_{1},\dotsc,j_{n}}=e_{j_{1}}\otimes\dotsb\otimes e_{j_{n}}$ as basis vectors of the space $V_{n}=V^{\otimes n}$ of $n$ qubits.
We also use the Pauli matrices $\{\sigma_{1},\sigma_{2},\sigma_{3}\}$ as a basis of simultaneously Hermitian and unitary operators acting on the space of each qubit, and set tensor products $\sigma_{j_{1},\dotsc,j_{n}}=\sigma_{j_{1}}\otimes\dotsb\otimes\sigma_{j_{n}}$ as basis operators acting on $V_{n}$. 

The GHZ state $u_{n}$ of $n$ qubits in its  non-normalized form is $u_{n}=e_{1,\dotsc,1}+e_{2,\dotsc,2}$.

The Dicke state $v_{n,m}$ of $n$ qubits of degree $m$ is a linear combinations of terms such that $e_{1}$ and $e_{2}$ appear $n-m$ and $m$ times, respectively, in each of them.
All possible ways to choose $m$ positions for $e_{2}$ among all possible $n$ positions contribute equally to the linear combination.
It follows that $v_{n,m}$ and $v_{n,n-m}$ are equivalent (by a change of basis) and thus it is sufficient to take $1\le m\le\floor{\frac{1}{2}n}$.
For example, the non-normalized Dicke states for $n=3$ and $n=4$ are
\begin{align}
    &v_{3,1}=e_{1,1,2}+e_{1,2,1}+e_{2,1,1}, \label{v_3_1} \\
    &v_{4,1}=e_{1,1,1,2}+e_{1,1,2,1}+e_{1,2,1,1}+e_{2,1,1,1}, \label{v_4_1} \\
    &v_{4,2}=e_{1,1,2,2}+e_{1,2,1,2}+e_{1,2,2,1}+e_{2,1,1,2}+e_{2,1,2,1}+e_{2,2,1,1}, \label{v_4_2}
\end{align}
where $v_{3,1}$ is also called the W state.

We note that the states $u_{n}$ are symmetric under the exchange of the basis elements $e_{1}$ and $e_{2}$ and that the states $v_{n,m}$ do not have such a symmetry for $n\not=2m$.
However, the method of eigenoperators developed in Section \ref{section_eigenoperators} requires the use of states with the above symmetry.
Consequently, we replace each state $v_{n,m}$ with the state $\tilde{v}_{n,m}$ that is the generalized symmetrization of $v_{n,m}$ under the exchange of the basis elements $e_{1}$ and $e_{2}$.
For $n=3$ and $n=4$ we define
\begin{align}
    &\tilde{v}_{3,1}=a_{1}(e_{2,1,1}+e_{1,2,2})+a_{2}(e_{1,2,1}+e_{2,1,2})+a_{3}(e_{1,1,2}+e_{2,2,1}), \label{tilde_v_3_1} \\
    &\tilde{v}_{4,1}=b_{1}(e_{2,1,1,1}+e_{1,2,2,2})+b_{2}(e_{1,2,1,1}+e_{2,1,2,2})+b_{3}(e_{1,1,2,1}+e_{2,2,1,2})+b_{4}(e_{1,1,1,2}+e_{2,2,2,1}), \label{tilde_v_4_1} \\
    &\tilde{v}_{4,2}=c_{1}(e_{1,1,2,2}+e_{2,2,1,1})+c_{2}(e_{1,2,1,2}+e_{2,1,2,1})+c_{3}(e_{1,2,2,1}+e_{2,1,1,2}), \label{tilde_v_4_2}
\end{align}
where $a_{i}$, $b_{i}$, $c_{i}$ are arbitrary non-zero complex numbers.

Entanglement properties of $v_{n,m}$ and $\tilde{v}_{n,m}$ are in general different, but in some cases they can be made identical.
For example, there is a change of basis that transforms $\tilde{v}_{3,1}$ into $v_{3,1}$ if and only if $a_{i}\not=\pm a_{j}$ for all $i\not=j$ and ${}\pm a_{1}\pm a_{2}\pm a_{3}=0$ for any one of the possible eight choices of signs.
This implies that the states $v_{3,1}$ and $\tilde{v}_{3,1}$ are essentially identical in these cases, and additionally, they belong to the same entanglement class in the algebraic entanglement classification method in \cite{Buniy:2010yh,Buniy:2010zp}. 
On the other hand, the states $v_{4,1}$ and $\tilde{v}_{4,1}$ for any $b_{i}$ belong to different entanglement classes and cannot be transformed into each other by a change of basis.
Finally, the states $v_{4,2}$ and $\tilde{v}_{4,2}$ belong to the same entanglement class and can be transformed into each other by a change of basis if and only if $\pm c_{1}=\pm c_{2}=\pm c_{3}$ for any one of the possible eight choices of signs.
 
We note that instead of the symmetry under the exchange of $e_{1}$ and $e_{2}$ (and the corresponding symmetry under the exchange of $\sigma_{1}$ and $\sigma_{2}$ in Section \ref{section_eigenoperators}), we could have used the corresponding anti-symmetry with similar results.

\section{Eigenoperators}
\label{section_eigenoperators}

We now define certain operators associated with the states $u_{n}$ and $\tilde{v}_{n,m}$.

We recall that commutativity of two operators is a sufficient (and not necessary) condition for the existence of a common eigenvector for the operators.
For a given state $v$, it is useful to find a set of $N(v)$ commuting eigenoperators $C(v)=\{C_{j}(v)\}_{1\le j\le N(v)}$, which are operators with a common eigenvector $v$.
Let $\gamma(v)=\{\gamma_{j}(v)\}_{1\le j\le N(v)}$ be the set of the corresponding eigenvalues, $C_{j}(v)v=\gamma_{j}(v)v$.
In addition to all elements of $C(v)$ commuting with each other, we restrict them to be linearly independent and thus $N(v)$ to be a finite number. 
We further restrict operators in $C(v)$ to those that are linear combinations of tensor products of only $\sigma_{1}$ and $\sigma_{2}$.
It follows from
\begin{align}
  &\sigma_{1}e_{1}=e_{2}, \ \sigma_{1}e_{2}=e_{1}, \ \sigma_{2}e_{1}=ie_{2}, \ \sigma_{2}e_{2}=-ie_{1}  \label{sigma_e}
\end{align}
that any state $v$ which is symmetric under permutations of $e_{1}$ and $e_{2}$ is an eigenvector with the eigenvalue either $1$ or $-1$ of an operator which is a tensor product of any number of $\sigma_{1}$s and an even number of $\sigma_{2}$s. 
This is the reason we have transformed non-symmetric states $v_{n,m}$ into symmetric states $\tilde{v}_{n,m}$ in Section \ref{section_states}.

With the above constraints on the form of operators $C_{j}(v)$, we solve the eigenvalue equations $C_{j}(v)v=\gamma_{j}(v)v$, obtain sets of commuting operators with common eigenvectors and the corresponding eigenvalues.
Table \ref{table_v_C_N_gamma} gives these operators and eigenvalues for the GHZ states $u_{n}$ and the generalized Dicke states $\tilde{v}_{n,m}$ of three and four qubits.
We note that $C(v)$ is a basis of eigenoperators for a given $v$ and any linear combination of them is also an eigenoperator for the same eigenvector $v$.

\begin{table}[htpb]
  \centering
  \begin{tabular}{cc}
    \toprule
    \multirow{3}{*}{$v$}
    & $C(v)$ \\
    & $N(v)$ \\
    & $\gamma(v)$ \\
    \midrule
    \multirow{3}{*}{$u_{3}$}
    & $\{\sigma_{1,1,1},\sigma_{1,2,2},\sigma_{2,1,2},\sigma_{2,2,1}\}$ \\
    & $\{\xi_{1}\xi_{2}\xi_{3},\xi_{1}\eta_{2}\eta_{3},\eta_{1}\xi_{2}\eta_{3},\eta_{1}\eta_{2}\xi_{3}\}$ \\
    & $\{1,-1,-1,-1\}$ \\
    \midrule
    \multirow{3}{*}{$\tilde{v}_{3,1}$}
    & $\{\sigma_{1,1,1},\tau_{3}\}$ \\
    & $\{\xi_{1}\xi_{2}\xi_{3},\mu_{3}\}$ \\
    & $\{1,1\}$ \\
    \midrule
    \multirow{3}{*}{$u_{4}$}
    & $\{\sigma_{1,1,1,1},\sigma_{1,1,2,2},\sigma_{1,2,1,2},\sigma_{1,2,2,1},\sigma_{2,1,1,2},\sigma_{2,1,2,1},\sigma_{2,2,1,1},\sigma_{2,2,2,2}\}$ \\
& $\{\xi_{1}\xi_{2}\xi_{3}\xi_{4},\xi_{1}\xi_{2}\eta_{3}\eta_{4},\xi_{1}\eta_{2}\xi_{3}\eta_{4},\xi_{1}\eta_{2}\eta_{3}\xi_{4},\eta_{1}\xi_{2}\xi_{3}\eta_{4},\eta_{1}\xi_{2}\eta_{3}\xi_{4},\eta_{1}\eta_{2}\xi_{3}\xi_{4},\eta_{1}\eta_{2}\eta_{3}\eta_{4}\}$ \\
    & $\{1,-1,-1,-1,-1,-1,-1,1\}$ \\
    \midrule
    \multirow{3}{*}{$\tilde{v}_{4,1}$}
    & $\{\sigma_{1,1,1,1},\tau_{4,1},\tau_{4,2},\tau_{4,3},\sigma_{2,2,2,2}\}$ \\
    & $\{\xi_{1}\xi_{2}\xi_{3}\xi_{4},\mu_{4,1},\mu_{4,2},\mu_{4,3},\eta_{1}\eta_{2}\eta_{3}\eta_{4}\}$ \\
    & $\{1,0,0,0,-1\}$ \\
    \midrule
    \multirow{3}{*}{$\tilde{v}_{4,2}$}
    & $\{\sigma_{1,1,1,1},\tau_{4,1,1},\tau_{4,1,2},\tau_{4,2,1},\tau_{4,2,2},\tau_{4,3,1},\tau_{4,3,2},\tau_{4,4,1},\tau_{4,4,2},\sigma_{2,2,2,2}\}$ \\
    & $\{\xi_{1}\xi_{2}\xi_{3}\xi_{4},\mu_{4,1,1},\mu_{4,1,2},\mu_{4,2,1},\mu_{4,2,2},\mu_{4,3,1},\mu_{4,3,2},\mu_{4,4,1},\mu_{4,4,2},\eta_{1}\eta_{2}\eta_{3}\eta_{4}\}$ \\
    & $\{1,1,1,1,1,1,1,1,1,1\}$ \\
    \bottomrule
 \end{tabular}
 \caption{Sets of commuting operators $C(v)$ with common eigenvectors $v$ and the corresponding sets of the instructional numbers $N(v)$ and eigenvalues $\gamma(v)$ for the GHZ states $u_{n}$ and the generalized Dicke states $\tilde{v}_{n,m}$ of three and four qubits.
     The sets $N(v)$ are obtained from the sets $C(v)$ by replacing the operators $\sigma_{1}$ and $\sigma_{2}$ acting on the $a$th qubit with the numbers $\xi_{a}$ and $\eta_{a}$ satisfying $\xi_{a},\eta_{a}\in\{-1,1\}$.
     The operators $\tau_{3},\tau_{4},\tau_{4,i},\tau_{4,i,j}$ and the numbers $\mu_{3},\mu_{4},\mu_{4,i},\mu_{4,i,j}$ are defined in \eqref{tau_3}, \eqref{tau_4}, \eqref{tau_4_1}, \eqref{tau_4_2}, \eqref{tau_4_3}, \eqref{tau_4_1_j}, \eqref{tau_4_2_j}, \eqref{tau_4_3_j}, \eqref{tau_4_4_j} and \eqref{mu_3}, \eqref{mu_4}, \eqref{mu_4_1}, \eqref{mu_4_2}, \eqref{mu_4_3}, \eqref{mu_4_1_j}, \eqref{mu_4_2_j}, \eqref{mu_4_3_j}, \eqref{mu_4_4_j}.
}
  \label{table_v_C_N_gamma}
\end{table}

We note that any linear combination of operators $\sigma_{j_{1},\dotsc,j_{n}}$ with an even number of indices $2$ has $u_{n}$ as its eigenvector.
However, only a much more restricted set of such linear combinations are eigenoperators for the states $\tilde{v}_{n,m}$.
For $n=3$ and $n=4$, these restricted linear combinations contain operators $\tau_{3}$ and $\tau_{4}$, $\tau_{4,i}$, $\tau_{4,i,j}$, respectively, where
\begin{align}
    &\tau_{3}=\sigma_{1,2,2}+\sigma_{2,1,2}+\sigma_{2,2,1}, \label{tau_3} \\
    &\tau_{4}=\sigma_{1,1,2,2}+\sigma_{1,2,1,2}+\sigma_{1,2,2,1}+\sigma_{2,1,1,2}+\sigma_{2,1,2,1}+\sigma_{2,2,1,1}, \label{tau_4} \\
    &\tau_{4,1}=\sigma_{1,1,2,2}+\sigma_{2,2,1,1}, \label{tau_4_1} \\
    &\tau_{4,2}=\sigma_{1,2,1,2}+\sigma_{2,1,2,1}, \label{tau_4_2} \\
    &\tau_{4,3}=\sigma_{1,2,2,1}+\sigma_{2,1,1,2}  \label{tau_4_3}
\end{align}
and
\begin{align}
    &\tau_{4,1,1}=\sigma_{1,1,2,2}+\sigma_{1,2,1,2}+\sigma_{1,2,2,1}, \quad \tau_{4,1,2}=\sigma_{2,1,1,2}+\sigma_{2,1,2,1}+\sigma_{2,2,1,1}, \label{tau_4_1_j} \\
    &\tau_{4,2,1}=\sigma_{1,1,2,2}+\sigma_{2,1,1,2}+\sigma_{2,1,2,1}, \quad \tau_{4,2,2}=\sigma_{1,2,1,2}+\sigma_{1,2,2,1}+\sigma_{2,2,1,1}, \label{tau_4_2_j} \\
    &\tau_{4,3,1}=\sigma_{1,2,1,2}+\sigma_{2,1,1,2}+\sigma_{2,2,1,1}, \quad \tau_{4,3,2}=\sigma_{1,1,2,2}+\sigma_{1,2,2,1}+\sigma_{2,1,2,1}, \label{tau_4_3_j} \\
    &\tau_{4,4,1}=\sigma_{1,2,2,1}+\sigma_{2,1,2,1}+\sigma_{2,2,1,1}, \quad \tau_{4,4,2}=\sigma_{1,1,2,2}+\sigma_{1,2,1,2}+\sigma_{2,1,1,2}. \label{tau_4_4_j}
\end{align}
We note that $\tau_{3}$, $\tau_{4}$, $\tau_{4,i}$ are non-factorizable and each $\tau_{4,i,j}$ is factorizable since it is the tensor product of $\sigma_{j}$ acting on the $i$th qubit and $\tau_{3}$ acting on all qubits other than the $i$th one.
These operators satisfy $\tau_{4,1}+\tau_{4,2}+\tau_{4,3}=\tau_{4}$ and $\tau_{4,i,1}+\tau_{4,i,2}=\tau_{4}$ for each $1\le i\le 4$, which implies that $\tau_{4}$ is also an eigenoperator for $\tilde{v}_{4,1}$ and $\tilde{v}_{4,2}$.

Although operators in each set $C(v)$ are linearly independent, there are non-linear relations between them that follow from the identities $\sigma_{1}^{2}=\sigma_{2}^{2}=I$ and $\sigma_{1}\sigma_{2}+\sigma_{2}\sigma_{1}=0$.
For example, for the operators in $C(u_{n})$, we have the relations
\begin{align}
    \sigma_{1,1,1}&=-\sigma_{1,2,2}\sigma_{2,1,2}\sigma_{2,2,1}, \label{identity_sigma_u_3} \\
    \sigma_{1,1,1,1}&=-\sigma_{1,1,2,2}\sigma_{1,2,1,2}\sigma_{1,2,2,1}=-\sigma_{1,1,2,2}\sigma_{2,1,1,2}\sigma_{2,1,2,1} \nn \\
    &=-\sigma_{1,2,1,2}\sigma_{2,1,1,2}\sigma_{2,2,1,1}=-\sigma_{1,2,2,1}\sigma_{2,1,2,1}\sigma_{2,2,1,1}, \label{identity_1_sigma_u_4} \\
    \sigma_{2,2,2,2}&=-\sigma_{2,1,1,2}\sigma_{2,1,2,1}\sigma_{2,2,1,1}=-\sigma_{1,2,1,2}\sigma_{1,2,2,1}\sigma_{2,2,1,1} \nn \\
    &=-\sigma_{1,1,2,2}\sigma_{1,2,2,1}\sigma_{2,1,2,1}=-\sigma_{1,1,2,2}\sigma_{1,2,1,2}\sigma_{2,1,1,2}. \label{identity_2_sigma_u_4}
\end{align}
The minus signs in \eqref{identity_sigma_u_3}, \eqref{identity_1_sigma_u_4}, \eqref{identity_2_sigma_u_4} are key for the analysis of Mermin devices for the GHZ states $u_{3}$ and $u_{4}$.

Non-linear relations between operators in $C(\tilde{v}_{n,m})$ are more complicated.
We seek them in the forms
\begin{align}
    &\sigma_{1,1,1}=f_{3}(\tau_{3}), \label{identity_sigma_tilde_v_3_1} \\
    &\sigma_{1,1,1,1}=f_{3}(\tau_{4,1,1})=f_{3}(\tau_{4,2,1})=f_{3}(\tau_{4,3,1})=f_{3}(\tau_{4,4,1}), \label{identity_1_sigma_tilde_v_4_1} \\
    &\sigma_{2,2,2,2}=f_{3}(\tau_{4,1,2})=f_{3}(\tau_{4,2,2})=f_{3}(\tau_{4,3,2})=f_{3}(\tau_{4,4,2}), \label{identity_2_sigma_tilde_v_4_1} \\
    &\sigma_{1,1,1,1}+\sigma_{2,2,2,2}=f_{4}(\tau_{4}), \label{identity_sigma_tilde_v_4_1}
\end{align}
where $f_{n}$ are certain polynomials.
The use of $f_{3}$ for $\tau_{4,i,j}$ reflects factorizability of $\tau_{4,i,j}$ in terms of $\tau_{3}$.
We look for the polynomials $f_{n}$ of the lowest order and with the smallest number of terms.
Each term in any even power of $\tau_{n}$ is the tensor product of an even number of the operators $\pm\sigma_{1}\sigma_{2}$ and the remaining number of the identity operators, and, consequently, these terms cannot produce the desired left-hand sides in \eqref{identity_sigma_tilde_v_3_1}, \eqref{identity_1_sigma_tilde_v_4_1}, \eqref{identity_2_sigma_tilde_v_4_1}, \eqref{identity_sigma_tilde_v_4_1}.
We are left with odd powers of $\tau_{n}$, and taking the smallest number of such powers of the lowest order, we find
\begin{align}
    &f_{3}(\tau)=\tfrac{1}{6}(-\tau^{3}+7\tau), \label{f_3} \\
    &f_{4}(\tau)=\tfrac{1}{24}(-\tau^{3}+28\tau). \label{f_4}
\end{align}

The operators $\tau_{4,i}$ satisfy the relations
\begin{align}
    &\sigma_{1,1,1,1}\sigma_{2,2,2,2}=-I+\tfrac{1}{2}\tau_{4,i}^{2}, \label{} \\
    &\sigma_{1,1,1,1}+\sigma_{2,2,2,2}=-\tfrac{1}{4}\tau_{4,1}\tau_{4,2}\tau_{4,3}. \label{}
\end{align}
In addition, $\tau_{4,i}^{3}=4\tau_{4,i}$, which implies $f_{3}(\tau_{4,i})=\frac{1}{2}\tau_{4,i}$ and $f_{4}(\tau_{4,i})=\tau_{4,i}$.

Properties of the operators $\tau_{3}$, $\tau_{4}$, $\tau_{4,i}$ and $\tau_{4,i,j}$ play a key role in the design and operation of Mermin devices for GHZ and generalized Dicke states to which we now turn.

\section{Mermin devices}
\label{section_mermin_devices}

Mermin devices show how quantum experiments prohibit something that classical experiments require.
We design such devices by proceeding similarly to the design in \cite{Mermin:1990AJP} for the GHZ state of three qubits.
In such a device, $n$ spin-$\frac{1}{2}$ particles are emitted from a source in a certain quantum state and arrive at the detector consisting of $n$ sub-detectors.
The detector is designed to measure expected values of quantum operators on states describing incoming particles.
For a given state of the particles $v$, we choose these operators from the set of commuting operators $C(v)$ for which $v$ is a common eigenvector. 
As a result, all measurements produce the set of the corresponding eigenvalues $\gamma(v)$ of the operators associated with the eigenvector $v$.

In an attempt to describe the results of the above experiments by classical means, we introduce Mermin's instructional sets such that a particle labeled by the index $a$ arriving from the source carries with it the instructional set $(\xi_{a},\eta_{a})$, where $\xi_{a},\eta_{a}\in\{-1,1\}$.
We replace every instance of the operators $\sigma_{1}$ and $\sigma_{2}$ acting on the $a$th qubit with the numbers $\xi_{a}$ and $\eta_{a}$, respectively.
This transforms the set of commuting operators $C(v)$ in Table \ref{table_v_C_N_gamma} into the corresponding sets of instructional numbers $N(v)$, which are also given in the table.
In particular the operators $\tau_{3}$, $\tau_{4}$, $\tau_{4,i}$, $\tau_{4,i,j}$ are replaced with the respective numbers $\mu_{3}$, $\mu_{4}$, $\mu_{4,i}$, $\mu_{4,i,j}$, where
\begin{align}
    &\mu_{3}=\xi_{1}\eta_{2}\eta_{3}+\eta_{1}\xi_{2}\eta_{3}+\eta_{1}\eta_{2}\xi_{3}, \label{mu_3} \\
    &\mu_{4}=\xi_{1}\xi_{2}\eta_{3}\eta_{4}+\xi_{1}\eta_{2}\xi_{3}\eta_{4}+\xi_{1}\eta_{2}\eta_{3}\xi_{4}+\eta_{1}\xi_{2}\xi_{3}\eta_{4}+\eta_{1}\xi_{2}\eta_{3}\xi_{4}+\eta_{1}\eta_{2}\xi_{3}\xi_{4}, \label{mu_4} \\
    &\mu_{4,1}=\xi_{1}\xi_{2}\eta_{3}\eta_{4}+\eta_{1}\eta_{2}\xi_{3}\xi_{4}, \label{mu_4_1} \\
    &\mu_{4,2}=\xi_{1}\eta_{2}\xi_{3}\eta_{4}+\eta_{1}\xi_{2}\eta_{3}\xi_{4}, \label{mu_4_2} \\
    &\mu_{4,3}=\xi_{1}\eta_{2}\eta_{3}\xi_{4}+\eta_{1}\xi_{2}\xi_{3}\eta_{4} \label{mu_4_3}
\end{align}
and
\begin{align}
    &\mu_{4,1,1}=\xi_{1}(\xi_{2}\eta_{3}\eta_{4}+\eta_{2}\xi_{3}\eta_{4}+\eta_{2}\eta_{3}\xi_{4}), \quad \mu_{4,1,2}=\eta_{1}(\xi_{2}\xi_{3}\eta_{4}+\xi_{2}\eta_{3}\xi_{4}+\eta_{2}\xi_{3}\xi_{4}), \label{mu_4_1_j} \\
    &\mu_{4,2,1}=\xi_{2}(\xi_{1}\eta_{3}\eta_{4}+\eta_{1}\xi_{3}\eta_{4}+\eta_{1}\eta_{3}\xi_{4}), \quad \mu_{4,2,2}=\eta_{2}(\xi_{1}\xi_{3}\eta_{4}+\xi_{1}\eta_{3}\xi_{4}+\eta_{1}\xi_{3}\xi_{4}), \label{mu_4_2_j} \\
    &\mu_{4,3,1}=\xi_{3}(\xi_{1}\eta_{2}\eta_{4}+\eta_{1}\xi_{2}\eta_{4}+\eta_{1}\eta_{2}\xi_{4}), \quad \mu_{4,3,2}=\eta_{3}(\xi_{1}\xi_{2}\eta_{4}+\xi_{1}\eta_{2}\xi_{4}+\eta_{1}\xi_{2}\xi_{4}), \label{mu_4_3_j} \\
&\mu_{4,4,1}=\xi_{4}(\xi_{1}\eta_{2}\eta_{3}+\eta_{1}\xi_{2}\eta_{3}+\eta_{1}\eta_{2}\xi_{3}), \quad \mu_{4,4,2}=\eta_{4}(\xi_{1}\xi_{2}\eta_{3}+\xi_{1}\eta_{2}\xi_{3}+\eta_{1}\xi_{2}\xi_{3}). \label{mu_4_4_j}
\end{align}

Experiments for a state $v$ with device configurations given by the operators in $C(v)$ can be explained with instructional sets if the equation
\begin{align}
    &N(v)=\gamma(v) \label{N_gamma}
\end{align}
is satisfied for some choice of $\xi_{a},\eta_{a}\in\{-1,1\}$.
In an equivalent description given in \cite{Mermin:1990AJP}, each sub-detector emits either red or green light after a measurement, where the color is determined by the configuration of a sub-detector and the instructional set of the particle that it absorbs. 
The specific instructions are that the variables $\xi_{a}$ and $\eta_{a}$ apply to sub-detectors in the configurations $\sigma_{1}$ and $\sigma_{2}$, respectively.
When the corresponding variable has the value $1$ or $-1$, the $a$th detector emits the red or green light, respectively.
The experiments can be described by instructional sets if there are no contradictions between numbers of lights of each color that sub-detectors emit.

We now check if description of the operation of the Mermin devices by instructional sets are possible for the GHZ and generalized Dicke states of three and four qubits. 

\subsection{GHZ states}
\label{section_ghz_states}

Since all eigenvalues in $\gamma(u_{n})$ for the GHZ states $u_{n}$ are either $-1$ or $1$, we choose the device to emit the red light for the eigenvalue $1$ and the green light for the eigenvalue $-1$.

We start with the state $u_{3}$.
We set up four experiments specified by the operators in $C(u_{3})$.
It follows from \eqref{N_gamma} and Table \ref{table_v_C_N_gamma} that we need to solve the system of equations
\begin{align}
    &\xi_{1}\xi_{2}\xi_{3}=1, \quad \xi_{1}\eta_{2}\eta_{3}=-1, \quad \eta_{1}\xi_{2}\eta_{3}=-1, \quad \eta_{1}\eta_{2}\xi_{3}=-1. \label{instructional_sets_equations_u_3}
\end{align}
Among all $64$ possible instructional sets, only $8$ sets satisfy the last three equations
in \eqref{instructional_sets_equations_u_3}.
However, none of these solutions satisfy the first equation in \eqref{instructional_sets_equations_u_3} because they all require $\xi_{1}\xi_{2}\xi_{3}=-1$.
Equivalently, multiplying all equations in \eqref{instructional_sets_equations_u_3}, we find $(\xi_{1}\xi_{2}\xi_{3}\eta_{1}\eta_{2}\eta_{3})^{2}=-1$, which is impossible. 
As a result, while the experiments with the device configurations $\sigma_{1,2,2}$, $\sigma_{2,1,2}$, $\sigma_{2,2,1}$ can be explained with instructional sets, no such explanation is possible when an additional experiment with the configuration $\sigma_{1,1,1}$ is added.
The above contradiction is solely due to the minus sign in the identity \eqref{identity_sigma_u_3}.
An equivalent formulation of the contradiction is to note that the numbers of red and green lights for the first experiment in \eqref{instructional_sets_equations_u_3} are odd and even, respectively, while for the last three experiments in \eqref{instructional_sets_equations_u_3} they are even and odd, respectively.

This is how Mermin originally showed that local instructional sets cannot explain all four experiments for the GHZ state $u_{3}$, but we note that Mermin used the state $2^{-1/2}(e_{1,1,1}-e_{2,2,2})$ instead of our state $u_{3}=e_{1,1,1}+e_{2,2,2}$.
We correspondingly switched the signs of the eigenvalues in $\gamma(u_{3})$ and the resulting parities of red and green lights.
Of course, the normalization plays no role in the above analysis.

The situation with the state $u_{4}$ is similar.
However, here we consider only subsets of operators in $C(u_{4})$ and the corresponding subsets in $N(u_{4})$ and $\gamma(u_{4})$.
There are four operators in each subset and there are eight ways to select them.
The operators are $\sigma_{1,1,1,1}$ (or $\sigma_{2,2,2,2}$, respectively) and one of the four sets of three operators appearing on the right hand side of \eqref{identity_1_sigma_u_4} (or \eqref{identity_2_sigma_u_4}, respectively).
It follows from \eqref{N_gamma} and Table \ref{table_v_C_N_gamma} (restricted to the selected subsets) that we need to solve each of the following systems of equations,
\begin{align}
    &\xi_{1}\xi_{2}\xi_{3}\xi_{4}=1, \quad \xi_{1}\xi_{2}\eta_{3}\eta_{4}=-1, \quad \xi_{1}\eta_{2}\xi_{3}\eta_{4}=-1, \quad \xi_{1}\eta_{2}\eta_{3}\xi_{4}=-1, \label{instructional_sets_equations_1_u_4} \\
    &\eta_{1}\eta_{2}\eta_{3}\eta_{4}=1, \quad \eta_{1}\xi_{2}\xi_{3}\eta_{4}=-1, \quad \eta_{1}\xi_{2}\eta_{3}\xi_{4}=-1, \quad \eta_{1}\eta_{2}\xi_{3}\xi_{4}=-1, \label{instructional_sets_equations_2_u_4} \\
    &\xi_{1}\xi_{2}\xi_{3}\xi_{4}=1, \quad \xi_{1}\xi_{2}\eta_{3}\eta_{4}=-1, \quad \eta_{1}\xi_{2}\xi_{3}\eta_{4}=-1, \quad \eta_{1}\xi_{2}\eta_{3}\xi_{4}=-1, \label{instructional_sets_equations_3_u_4} \\
    &\eta_{1}\eta_{2}\eta_{3}\eta_{4}=1, \quad \xi_{1}\eta_{2}\xi_{3}\eta_{4}=-1, \quad \xi_{1}\eta_{2}\eta_{3}\xi_{4}=-1, \quad \eta_{1}\eta_{2}\xi_{3}\xi_{4}=-1, \label{instructional_sets_equations_4_u_4} \\
    &\xi_{1}\xi_{2}\xi_{3}\xi_{4}=1, \quad \xi_{1}\eta_{2}\xi_{3}\eta_{4}=-1, \quad \eta_{1}\xi_{2}\xi_{3}\eta_{4}=-1, \quad \eta_{1}\eta_{2}\xi_{3}\xi_{4}=-1, \label{instructional_sets_equations_5_u_4} \\
    &\eta_{1}\eta_{2}\eta_{3}\eta_{4}=1, \quad \xi_{1}\xi_{2}\eta_{3}\eta_{4}=-1, \quad \xi_{1}\eta_{2}\eta_{3}\xi_{4}=-1, \quad \eta_{1}\xi_{2}\eta_{3}\xi_{4}=-1, \label{instructional_sets_equations_6_u_4} \\
    &\xi_{1}\xi_{2}\xi_{3}\xi_{4}=1, \quad \xi_{1}\eta_{2}\eta_{3}\xi_{4}=-1, \quad \eta_{1}\xi_{2}\eta_{3}\xi_{4}=-1, \quad \eta_{1}\eta_{2}\xi_{3}\xi_{4}=-1, \label{instructional_sets_equations_7_u_4} \\
    &\eta_{1}\eta_{2}\eta_{3}\eta_{4}=1, \quad \xi_{1}\xi_{2}\eta_{3}\eta_{4}=-1, \quad \xi_{1}\eta_{2}\xi_{3}\eta_{4}=-1, \quad \eta_{1}\xi_{2}\xi_{3}\eta_{4}=-1. \label{instructional_sets_equations_8_u_4} 
\end{align}
Multiplying all equations in each of the systems \eqref{instructional_sets_equations_1_u_4}, \eqref{instructional_sets_equations_2_u_4}, \eqref{instructional_sets_equations_3_u_4}, \eqref{instructional_sets_equations_4_u_4}, \eqref{instructional_sets_equations_5_u_4}, \eqref{instructional_sets_equations_6_u_4}, \eqref{instructional_sets_equations_7_u_4}, \eqref{instructional_sets_equations_8_u_4}, we find that each of the resulting eight equations has the form of the product of even powers of the variables $\xi_{a}$ and $\eta_{a}$ being equal to $-1$, which is impossible. 
As a result, while the experiments with the device configurations given by the operators on the right hand side of \eqref{identity_1_sigma_u_4} (or \eqref{identity_2_sigma_u_4}, respectively) can be explained by instructional sets, no such explanation is possible when an additional experiment with the configuration $\sigma_{1,1,1,1}$ (or $\sigma_{2,2,2,2}$, respectively) is added.

We note that multiplying one of the identities in \eqref{identity_1_sigma_u_4} and another one in \eqref{identity_2_sigma_u_4}, we obtain
\begin{align}
    &\sigma_{1,1,1,1}\sigma_{2,2,2,2}=\sigma_{1,1,2,2}\sigma_{1,2,1,2}\sigma_{1,2,2,1}\sigma_{2,1,1,2}\sigma_{2,1,2,1}\sigma_{2,2,1,1}. \label{square_identity_u_4}
\end{align}
The absence of the minus sign in \eqref{square_identity_u_4} is the reason for not setting up the Mermin device for $u_{4}$ to measure all operators in $C(u_{4})$ instead of only their subsets; if we have done so, no contradiction would resulted from the use of instructional sets.

\subsection{Generalized Dicke states}
\label{section_generalized_dicke_states}

Construction of Mermin devices for the generalized Dicke states $\tilde{v}_{n,m}$ differ from the construction for the GHZ states $u_{n}$ only in the choice of operators that the devices measure.
For the state $\tilde{v}_{3,1}$, we do experiments with $\sigma_{1,1,1}$ and $\tau_{3}$ and according to \eqref{N_gamma} and Table \ref{table_v_C_N_gamma}, we need to solve the system of equations
\begin{align}
    &\xi_{1}\xi_{2}\xi_{3}=1, \quad \mu_{3}=1. \label{instructional_sets_equations_v_3_1}
\end{align}
There are $24$ solutions of the second equation in \eqref{instructional_sets_equations_v_3_1},  but they all give $\xi_{1}\xi_{2}\xi_{3}=-1$.
We also note that the identity \eqref{identity_sigma_tilde_v_3_1} implies $f_{3}(\mu_{3})=1$, and the only solutions of this equation are $\mu_{3}\in\{-3,1,2\}$.
We check that in addition to the above $24$ solutions for $\mu_{3}=1$, there are also eight solutions for $\mu_{3}=-3$ (for all of which $\xi_{1}\xi_{2}\xi_{3}=-1$ again) and no solutions for $\mu_{3}=2$.
We thus conclude that instructional sets cannot describe measurements of $\sigma_{1,1,1}$ and $\tau_{3}$ made by the Mermin device for the state $\tilde{v}_{3,1}$. 

For the state $\tilde{v}_{4,1}$, we solve the system of equations
\begin{align}
    &\xi_{1}\xi_{2}\xi_{3}\xi_{4}=1, \quad \mu_{4,1}=\mu_{4,2}=\mu_{4,3}=0, \quad \eta_{1}\eta_{2}\eta_{3}\eta_{4}=-1 \label{instructional_sets_equations_v_4_1}
\end{align}
to find $64$ solutions for $\xi_{a},\eta_{a}\in\{-1,1\}$.
It might appear that this conclusion depends on a particular choice that we made in Table \ref{table_v_C_N_gamma} for the set $C(\tilde{v}_{4,1})$.
However, it can be shown that that particular set is a basis of all eigenoperators with real eigenvalues for the eigenvector $\tilde{v}_{4,1}$.
Using any linear combinations of the operators from this basis leads to similar solutions. 
We thus conclude that instructional sets (somewhat surprisingly) can describe measurements of the Mermin device for the state $\tilde{v}_{4,1}$. 

Finally, for the state $\tilde{v}_{4,2}$, it is sufficient to choose the operators $\sigma_{1,1,1,1}$, $\sigma_{2,2,2,2}$ and one of the operators $\tau_{4,i,j}$ from the set $C(\tilde{v}_{4,2})$, in which case we have eight systems of three equations
\begin{align}
    &\xi_{1}\xi_{2}\xi_{3}\xi_{4}=1, \quad \mu_{4,i,j}=1, \quad \eta_{1}\eta_{2}\eta_{3}\eta_{4}=1. \label{instructional_sets_equations_v_4_2}
\end{align}
Every one of these systems is incompatible with the requirement $\xi_{a},\eta_{a}\in\{-1,1\}$.
Including more than one operator $\tau_{4,i,j}$ leads to the same negative result.
We thus conclude that instructional sets cannot describe measurements of the Mermin device for the state $\tilde{v}_{4,2}$. 
The above results deserve further studies and we will report on their generalizations to larger number of qubits in a forthcoming work \cite{BK2026}.

For an earlier version of a Mermin device for the W state of $3$ qubits, see \cite{pitowsky1991relativity}.

\section{Bounds}
\label{section_bounds}

We look at the commuting sets of operators $C(u_{n})$ and $C(\tilde{v}_{n,m})$ listed in Table \ref{table_v_C_N_gamma}.
As the Pauli matrices $\sigma_{j}$ are simultaneously Hermitian and unitary (so that $\sigma_{j}^{2}=I$), we replace the operators $(\sigma_{1},\sigma_{2})$ acting on the $a$th qubit with the operators $(X_{a},Y_{a})$.
Here each operator $X_{a}$ and $Y_{a}$ is simultaneously Hermitian and unitary (so that $X_{a}^{2}=Y_{a}^{2}=I$).
After this replacement, we take into consideration that any linear combination of a set of operators with a common eigenvector has the same eigenvector.
We are consequently led to consider the Mermin operators $M_{n}$, where, for example,
\begin{align}
    M_{3}&=X_{1}\otimes X_{2}\otimes X_{3}-X_{1}\otimes Y_{2}\otimes Y_{3}-Y_{1}\otimes X_{2}\otimes Y_{3}-Y_{1}\otimes Y_{2}\otimes X_{3}, \label{M_3} \\
    M_{4}&=X_{1}\otimes X_{2}\otimes X_{3}\otimes X_{4}-X_{1}\otimes X_{2}\otimes Y_{3}\otimes Y_{4}-X_{1}\otimes Y_{2}\otimes X_{3}\otimes Y_{4}-X_{1}\otimes Y_{2}\otimes Y_{3}\otimes X_{4} \nn \\
    &-Y_{1}\otimes X_{2}\otimes X_{3}\otimes Y_{4}-Y_{1}\otimes X_{2}\otimes Y_{3}\otimes X_{4}-Y_{1}\otimes Y_{2}\otimes X_{3}\otimes X_{4}+Y_{1}\otimes Y_{2}\otimes Y_{3}\otimes Y_{4}. \label{M_4}
\end{align}
It follows from $X_{a}^{2}=Y_{a}^{2}=I$ that the norms of $X_{a}$ and $Y_{a}$ satisfy $\smallnorm{X_{a}}=\smallnorm{Y_{a}}=1$, and now \eqref{M_3} and \eqref{M_4} lead to $\smallnorm{M_{3}}\le 4$, $\smallnorm{M_{4}}\le 8$.

For any state $v\in V_{n}$, we define the expected value $\mu_{n}(v)=\smallnorm{v}^{-2}\smallproduct{v}{M_{n}v}$, and note that $\smallabs{\mu_{n}(v)}\le\smallnorm{M_{n}}$.
If $v$ is an eigenvector of $M_{n}$, then $\mu_{n}(v)$ equals the corresponding eigenvalue.
The choice of the operators \eqref{M_3} and \eqref{M_4} is natural from the point of view that for the restricted case $X_{a}=\sigma_{1}$ and $Y_{a}=\sigma_{2}$ for all $1\le a\le n$ the corresponding GHZ states $u_{3}$ and $u_{4}$ are their respective eigenvectors, $M_{3}u_{3}=4u_{3}$ and $M_{4}u_{4}=8u_{4}$.
The corresponding expected values $\mu_{3}(u_{3})=4$ and $\mu_{4}(u_{4})=8$ saturate the bounds $\smallabs{\mu_{n}(u_{n})}\le\smallnorm{M_{n}}$.
We also note that, for the above restricted case, the generalized Dicke states $\tilde{v}_{3,1}$, $\tilde{v}_{4,1}$, $\tilde{v}_{4,2}$ are eigenvectors of the above operators with zero eigenvalues, $M_{3}\tilde{v}_{3,1}=0$, $M_{4}\tilde{v}_{4,1}=0$, $M_{4}\tilde{v}_{4,2}=0$.
This leads to $\mu_{3}(\tilde{v}_{3,1})=\mu_{4}(\tilde{v}_{4,1})=\mu_{4}(\tilde{v}_{4,2})=0$, in contrast to the maximal property of the GHZ states $u_{3}$ and $u_{4}$.

For a general case of the operator $M_{n}$, we choose it from the set of all operators $\mathcal{M}_{n}$ obtained by taking arbitrary $X_{a}$ and $Y_{a}$ in \eqref{M_3} and \eqref{M_4} that satisfy $X_{a}^{2}=Y_{a}^{2}=I$.
We consequently define $m_{n}(v)=\max{\{\smallabs{\mu_{n}(v)}\colon M_{n}\in\mathcal{M}_{n}\}}$.
This maximum exists in view of the bound $\smallabs{\mu_{n}(v)}\le\smallnorm{M_{n}}$.
Analytic calculations for the GHZ and Dicke states of three and four qubits give
\begin{align}
    &m_{3}(u_{3})=4, \label{m_3_u_3} \\
    &m_{3}(v_{3,1})=c=\frac{1}{9}\sqrt{738\sqrt{41}-3974}\approx 3.04596, \label{m_3_v_3_1} \\
    &m_{4}(u_{4})=8, \label{m_4_u_4} \\
    &m_{4}(v_{4,1})=\tfrac{9}{2}, \label{m_4_v_4_1} \\
    &m_{4}(v_{4,2})=6. \label{m_4_v_4_2}
\end{align}

The results \eqref{m_3_u_3} and \eqref{m_4_u_4} for the GHZ states are well known.
The approximate numerical value for $m_{3}(v_{3,1})$ can be found in the literature, but to our knowledge, not the exact analytic expression \eqref{m_3_v_3_1}.
Further analytic and numerical results can be found in \cite{BK2026}.

We now proceed with the derivations of the results \eqref{m_3_v_3_1}, \eqref{m_4_v_4_1}, \eqref{m_4_v_4_2} for the Dicke states.
It turns out that these derivations can be done for restricted forms of the operators \eqref{M_3} and \eqref{M_4} that have $X_{a}=X$ and $Y_{a}=Y$ for all $1\le a\le n$, where $X$ and $Y$ are simultaneously Hermitian and unitary (and thus satisfying $X^{2}=Y^{2}=I$).
We expand $X=\sum_{j=1}^{3}x_{j}\sigma_{j}$, $Y=\sum_{j=1}^{3}y_{j}\sigma_{j}$, where $x_{j},y_{j}\in\R$ and $\sigma_{j}$ are the Pauli matrices. 
The conditions $X^{2}=Y^{2}=I$ become $\smallnorm{x}=\smallnorm{y}=1$.

Direct calculations of the expected values of the Mermin operators \eqref{M_3} and \eqref{M_4} for the Dicke states of three and four qubits give
\begin{align}
    \mu_{3}(v_{3,1})&=-3x_{3}^{3}+5x_{3}y_{3}^{2}-4(x_{1}y_{1}+x_{2}y_{2})y_{3}, \label{mu_3_v_3_1} \\
    \mu_{4}(v_{4,1})&=-4(x_{3}^{4}+y_{3}^{4})+12x_{3}^{2}y_{3}^{2}-12(x_{1}y_{1}+x_{2}y_{2})x_{3}y_{3}, \label{mu_4_v_4_1} \\
    \mu_{4}(v_{4,2})&=6(x_{3}^{4}+y_{3}^{4})-16x_{3}^{2}y_{3}^{2}-4(x_{1}y_{1}+x_{2}y_{2})^{2}+16(x_{1}y_{1}+x_{2}y_{2})x_{3}y_{3}. \label{mu_4_v_4_2}
\end{align}
It can be shown that the global extremums of \eqref{mu_3_v_3_1}, \eqref{mu_4_v_4_1}, \eqref{mu_4_v_4_2} occur when $\smallabs{x_{1}y_{1}+x_{2}y_{2}}$ takes its largest value. 
Since $x$ and $y$ are unit vectors, it follows that this corresponds to
\begin{align}
    &x_{1}y_{1}+x_{2}y_{2}=\pm(1-x_{3}^{2})^{1/2}(1-y_{3}^{2})^{1/2}. \label{x_1_x_2_y_1_y_2}
\end{align}
Geometrically, \eqref{x_1_x_2_y_1_y_2} means that the projections of $x$ and $y$ on the $(1,2)$ plane are collinear.
It follows that
\begin{align}
    &x_{1}=(1-x_{3}^{2})^{1/2}\cos{\phi}, \quad x_{2}=(1-x_{3}^{2})^{1/2}\sin{\phi}, \label{x_answer} \\
    &y_{1}=(1-y_{3}^{2})^{1/2}\cos{\psi}, \quad y_{2}=(1-y_{3}^{2})^{1/2}\sin{\psi}, \label{y_answer}
\end{align}
where arbitrary $\phi$ and $\psi$ are constrained by $\cos{(\phi-\psi)}=\pm 1$.

\begin{figure}[htpb]
    \centering
    \includegraphics[width=\textwidth]{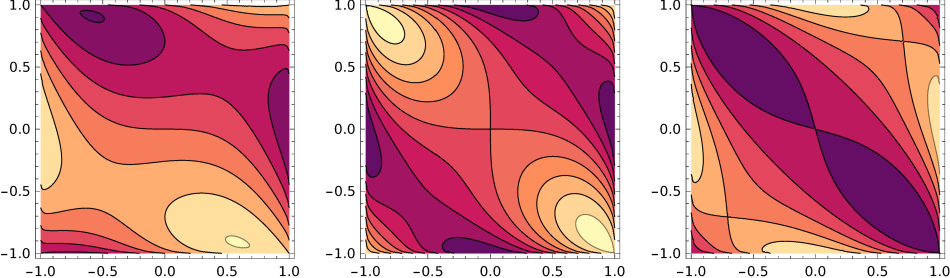}
    \caption{Contour plots of $\mu_{3,+}(v_{3,1})$, $\mu_{4,+}(v_{4,1})$ and $\mu_{4,+}(v_{4,2})$ as functions of $x_{3}$ and $y_{3}$.
	Extremal values in \eqref{m_3_v_3_1}, \eqref{m_4_v_4_1} and \eqref{m_4_v_4_2} are at the locations in these figures that are given by the coordinates $(x_{3},y_{3})$ in \eqref{x_3_y_3_v_3_1}, \eqref{x_3_y_3_v_4_1} and \eqref{x_3_y_3_v_4_2}.
}
    \label{figure_contours}
\end{figure}

Substituting \eqref{x_1_x_2_y_1_y_2} into $\mu_{n}(v_{n,m})$ in \eqref{mu_3_v_3_1}, \eqref{mu_4_v_4_1}, \eqref{mu_4_v_4_2}, we obtain the functions $\mu_{n,\pm}(v_{n,m})$ that depend only on $x_{3}$ and $y_{3}$, where the $\pm$ sign in $\mu_{n,\pm}$ correspond to the $\pm$ sign in \eqref{x_1_x_2_y_1_y_2}. 
The contour plots of the functions $\mu_{n,+}(v_{n,m})$ are given in Figure \ref{figure_contours} and the corresponding plots of the functions $\mu_{n,-}(v_{n,m})$ are obtained by the reflection with respect to the line $y_{3}=0$.
In all the above cases, extremums of $\mu_{n,\pm}(v_{n,m})$ can be found analytically and the corresponding results for the quantities $m_{n}(v_{n,m})=\max{\smallabs{\mu_{n}(v_{n,m})}}$ are given in \eqref{m_3_v_3_1}, \eqref{m_4_v_4_1}, \eqref{m_4_v_4_2} and the required values of $x_{3}$ and $y_{3}$ are 
\begin{align}
    &(x_{3},y_{3})=(sa,tb), \quad a=\frac{\sqrt{3\sqrt{41}-13}}{3\sqrt{2}}\approx 0.587337, \quad b=\frac{\sqrt{5\sqrt{41}-27}}{\sqrt{6}}\approx 0.914296 \ \  \text{for\ } v_{3,1}, \label{x_3_y_3_v_3_1} \\
    &(x_{3},y_{3})=\tfrac{1}{2}\sqrt{3}(s,t) \ \  \text{for\ } v_{4,1}, \label{x_3_y_3_v_4_1} \\
    &(x_{3},y_{3})\in\{(s,0),(0,t),2^{-1/2}(s,t)\} \ \  \text{for\ } v_{4,2}, \label{x_3_y_3_v_4_2}
\end{align}
where all four possible choice for $s,t\in\{-1,1\}$ are allowed.

\section{Conclusions}
\label{section_conclusions}

Mermin devices allow deeper and simpler understanding of impossibility of description of quantum experiments in terms of local hidden variables.
Such devices are based on existence of sets of commuting operators and certain identities among them that lead to contradictions when operators are replaced with local instructional sets.
We have generalized this approach from the original Mermin construction for the GHZ state of three qubits to the GHZ state of four qubits as well as the generalized Dicke states of three and four qubits.
This requires more general approach when choosing appropriate sets of commuting operators for these states.

We also presented exact results for bounds of expected values of Mermin operators for the above states that describe degrees of violation of the Bell inequalities for them.
In particular, our analytical result for the W state of three qubits (that, to our knowledge, does not exist in the literature) agrees with the known numerical results.
Similarly to maximum violation of the Bell inequalities by the GHZ states, the analogous results for the W and Dicke states give us more detailed description of quantumness of these states.

We will generalize and extend these results further in an upcoming work that considers a broader class of states of larger numbers of qubits \cite{BK2026} .
Ideally, we would like to include analogous description for all entanglement classes of states of four qubits, as in our classification in \cite{Buniy:2010yh,Buniy:2010zp}.



\end{document}